\documentclass[12pt]{article}
\usepackage{latexsym,amssymb,amstext}
\newcommand{\be}{\begin{equation}}
\newcommand{\ee}{\end{equation}}
\newcommand{\beqs}{\begin{eqnarray}}
\newcommand{\eeqs}{\end{eqnarray}}

\def\({\left(}
\def\){\right)}

\newcommand{\Exc}[1]{{${\rm E}_{{#1}({#1})}$}}

\def\mxth{\mathsurround=0pt }
\def\xversim#1#2{\lower2.pt\vbox{\baselineskip0pt \lineskip-.5pt
x  \ialign{$\mxth#1\hfil##\hfil$\crcr#2\crcr\sim\crcr}}}

\newcommand{\pa}{\partial}

\newcommand{\m}{\mu}
\newcommand{\n}{\nu}

\def\be{\begin{equation}}
\def\ee{\end{equation}}
\def\bea{\begin{eqnarray}}
\def\eea{\end{eqnarray}}

\newcommand{\ft}[2]{{\textstyle\frac{#1}{#2}}}

\def\bfone{\relax{\rm 1\kern-.35em 1}}

\makeatletter
\@addtoreset{equation}{section}
\makeatother

\begin{document}
\begin{titlepage}

\begin{center}
ITP-UU-05/21 \qquad\quad SPIN-05/16 \qquad\quad DESY 05-134 
\quad\quad
ZMP-HH/05-14
\end{center}
\vskip 22mm
\begin{center}
{\Large {\bf Magnetic charges in local field theory}}
\end{center}
\vskip 6mm

\begin{center}
{{\bf Bernard de Wit}\\
 Institute for Theoretical Physics \,\&\, Spinoza Institute,\\  
Utrecht University, Postbus 80.195, NL-3508 TD Utrecht, 
The Netherlands\\
{\tt b.dewit@phys.uu.nl}}
\vskip 2mm
{{\bf Henning Samtleben}\\
II. Institut f\"ur Theoretische Physik der Universit\"at Hamburg,\\ 
 Luruper Chaussee 149, D-22761 Hamburg, Germany\\
{\tt henning.samtleben@desy.de}}
\vskip 2mm
{{\bf Mario Trigiante}\\
Dept. of Physics, Politecnico di Torino,\\
Corso Duca degli Abruzzi 24, I-10129 Torino, Italy \\
{\tt mario.trigiante@to.infn.it}}
\vskip 4mm

\end{center}

\vskip .2in

\begin{center} {\bf Abstract } \end{center}
\begin{quotation}\noindent
  Novel Lagrangians are discussed in which (non-abelian) electric and
  magnetic gauge fields appear on a par. To ensure that these
  Lagrangians describe the correct number of degrees of freedom,
  tensor gauge fields are included with corresponding gauge
  symmetries.  Non-abelian gauge symmetries that involve both the
  electric and the magnetic gauge fields can then be realized at the
  level of a single gauge invariant Lagrangian, without the need of
  performing duality transformations prior to introducing the gauge
  couplings. The approach adopted, which was initially developed for
  gaugings of maximal supergravity, is particularly suited for the
  study of flux compactifications.
\end{quotation}
\end{titlepage}
\eject
\section{Introduction}
It is generally believed that electric and magnetic gauge fields
cannot be described in terms of a single local Lagrangian. While
electric and magnetic fields are defined by field strengths which can
be related in a local fashion, the underlying vector potentials, in
general, do not satisfy such a relation (this paper deals with
four-dimensional gauge theories, so that both electric and magnetic
potentials are vector fields). In certain cases, for instance, when
describing magnetic monopoles in terms of the electric vector
potential, the latter cannot be single-valued. In the absence of
charges, the Bianchi identities (which imply that the field strengths
can be expressed in terms of vector potentials) and the field
equations for these vector potentials take a similar form.  Assuming
that we are dealing with $n$ vector potentials, we have $n$ Bianchi
identities and $n$ field equations. Upon rotating these $2n$ equations
(by a symplectic $2n$-by-$2n$ matrix) there is the option of selecting
$n$ independent linear combinations of them to be interpreted as
Bianchi identities whose solutions lead to a different set of vector
potentials. In terms of these vector potentials there exists a
different Lagrangian which gives rise to field equations and Bianchi
identities for the field strengths that are linearly equivalent to the
original ones. However, the two dual sets of gauge fields are not
locally related and therefore they cannot appear simultaneously in a
given local Lagrangian. By the same token one cannot have a local
coupling of the gauge fields to magnetic charges as the latter couple
locally to the dual magnetic gauge fields.

Hence different Lagrangians can describe the same set of field
equations and Bianchi identities. The transformations governing these
inequivalent Lagrangians are known as electric/magnetic duality.
Symmetries of the combined equations of motion and Bianchi identities
are not necessarily realized at the level of the Lagrangian and may
involve a subgroup of the electric/magnetic duality transformations.
This poses a problem when switching on (possibly nonabelian) gauge
interactions, as the gauging must proceed through electric gauge
fields. A coupling to the magnetic charges seems therefore only
possible after applying an appropriate electric/magnetic duality
transformation by which all relevant charges are converted to electric
ones. In this paper we demonstrate in a rather general framework how
nevertheless one can have couplings to the magnetic charges without
first converting them to electric ones. Here we should stress that we
restrict ourselves to electric/magnetic charges that are mutually
local, so that problems of a more fundamental nature are avoided. 

The situation described above has an analogue in space-time
dimensions other than four, where electric/magnetic duality
takes the form of a duality between vector and tensor gauge fields. In
$d$ space-time dimensions, antisymmetric gauge fields of rank-$p$ are
dual to antisymmetric gauge fields of rank $d-2-p$. So the dual gauge
field of an electric vector potential is an antisymmetric gauge field
of rank $d-3$. While the former couples naturally to an electrically
charged particle, the latter couples to a magnetically charged brane
of $(d-4)$-dimensional spatial extension. When attempting to switch on
gauge interactions one encounters the same problem as noted above in
the four-dimensional context. Namely, one has to convert all the gauge 
fields that are supposed to couple to the charges to (electric) vector
potentials. Furthermore, because the remaining vector gauge fields
must be neutral, one must convert the charged vector fields that
do not belong to the adjoint representation of the gauge group, to
tensor fields. While this seems to pose no problem of principle, the field
content of the theory thus  depends sensitively on the gauging, so that
introducing a gauge group is no longer a matter of simply
switching on a corresponding gauge coupling constant. This fact
precludes any uniform treatment of the gaugings of these theories and
furthermore thoroughly obscures the symmetry structure of the
underlying ungauged theory. 

Recently, in our study of gaugings of maximal supergravities in five
space-time dimensions \cite{deWit:2004nw} we exploited a framework in
which these conversions between vector and tensor fields are no longer
necessary. Gaugings are encoded in a so-called embedding tensor
which defines the embedding of the gauge group into the symmetry group
of the ungauged theory. The symmetry structure of the latter remains
completely manifest (although the full symmetry is broken by the
embedding tensor) because both vector and tensor fields are present
and assigned to representations of the symmetry group of the ungauged
theory. The presence of an intricate set of vector-tensor gauge
transformations ensures that the number of physical degrees of freedom
remains the same. In \cite{deWit:2004nw} it was already noted how this
approach can be applied to gaugings of maximal supergravity in various
possible space-time dimensions. As was explained in~\cite{deWit:2005hv}, 
one is generically dealing with hierarchies of
vector-tensor gauge fields and there exists an intriguing interplay
between the group-theoretical assignment of the various tensor fields
and their duality relation.  Although this was primarily explained in
the context of the \Exc{k} duality groups of maximal supergravity, the
mechanism is by no means restricted to supergravity and can be applied
to generic gauge theories. This paper explains how this is done in the
context of four space-time dimensions, where the precise
implementation of the mechanism was yet unknown. 

As discussed in \cite{deWit:2005hv}, the gauge theory in four
space-time dimensions is augmented by rank-2 tensor fields
transforming in the adjoint representation of the symmetry group of
the ungauged theory (in this case \Exc7, as this is the symmetry group
of the ungauged maximal supergravity). The analysis was based on the
group-theoretical properties of the embedding tensor, which we will
have to determine in the more general setting, as shall be discussed
in due course. The presence of these tensor fields is reminiscent of a
similar situation in the context of $N=2$ supergravity. In
\cite{Louis:2002ny} it was noted that flux compactifications of
type-II theories on Calabi-Yau threefolds gave rise to a gauged $N=2$
supergravity with electric and magnetic charges, which was not of the
`canonical type'. The presence of the tensor field is by itself not
surprising in the context of a compactification from
higher-dimensional supergravity. Because the gauging was abelian its
effect was confined to the interactions of the tensor field with the
vector gauge fields (apart from a scalar potential required by
supersymmetry). In that particular case the fluxes are such that 
the theory remains symplectically invariant. The term symplectic
`invariance' is perhaps somewhat misleading. Rephrasing the latter
result in the context of our work \cite{deWit:2002vt,deWit:2004nw}, it
means that the embedding tensor is parametrized in terms of $2n$
charges, which, when treated as spurionic quantities, preserve the
manifest symplectic invariance.  This `spurionic' approach was a
crucial ingredient of the group-theoretical analysis of
\cite{deWit:2002vt,deWit:2004nw}. The symplectic invariance is thus an
equivalence relation between two theories, rather than an invariance. 

As we pointed out previously, we will be dealing with charges that are
mutually local so that they can be converted to electric ones.  In
principle there is nothing wrong in having to perform a series of
field dualities prior to switching on the charges. However, doing so
will always obscure the symmetry structure that the theory has
inherited from the ungauged theory. Moreover, the Lagrangian often
contains terms that diverge in the limit of vanishing gauge coupling
constant, and there are also a number of practical drawbacks. When
performing a symplectic reparametrization on the charges in the
compactification discussed in \cite{Louis:2002ny}, the theory can be
brought in the more `canonical' form that is known for $N=2$
supergravity
\cite{deWit:1984pk,deWit:1984px,DAuria:1990fj,DAuria:2001kv,deWit:2001bk}.  
However, the manifest
symplectic invariance is lost in that case. In the approach that we
discuss in this paper, all of this is no longer necessary and moreover
one can also discuss non-abelian gauge groups. In the formulation that
we will present there is a topological term that takes a universal
form encoded in terms of the embedding tensor and depending only on
the vector and tensor fields. The Lagrangian is fully gauge invariant,
irrespective of whether the original rigid invariance was respected by
the initial Lagrangian, or only by the combined field equations and
Bianchi identities. When imposing a gauge choice and integrating out
certain fields (which originate from other parts in the Lagrangian)
the universal features are lost and a large variety of Lagrangians is
generated.

This paper is organized as follows. In section~2 we briefly discuss
the issue of electric/magnetic duality. The gaugings with both
electric and magnetic vector potentials is introduced in section~3,
where we present the constraints on the embedding tensor and explain
the connection with the tensor fields transforming in the adjoint
representation of the symmetry group of the ungauged theory. In
section~4 we derive the universal gauge invariant interactions for the
vector-tensor system, which are fully encoded in terms of the
embedding tensor. In section~5 we elucidate some of our results. We
demonstrate how all tensor fields can be integrated out from the
Lagrangian in the presence of non-abelian gauge interactions, which
amounts to effecting an electric/magnetic duality
transformation in the presence of charges at the Lagrangian level.
We also discuss some features relevant to abelian gaugings and to the
application of our formalism to gauged $N=2$ supergravity.
Undoubtedly our result will have many other applications with or without supersymmetry, 
but we refrain from presenting further explicit examples here.

\section{Electric/magnetic duality}
In the absence of charges, gauge invariant Lagrangians in four
space-time dimensions based on abelian gauge fields $A_\mu{}^\Lambda$, 
labeled by the index $\Lambda=1,\ldots,n$, can be expressed in terms of
their abelian field strengths, $\mathcal{F}_{\mu\nu}{}^\Lambda =
2\,\partial_{[\mu}A_{\nu]}{}^\Lambda$. The field equations for these
fields and the Bianchi identities for the field strengths constitute
$2n$ equations,
\begin{equation}
  \label{eq:eom-bianchi}
  \partial_{[\mu} \mathcal{F}_{\nu\rho]}{}^\Lambda  
   = 0 = \partial_{[\mu} \mathcal{G}_{\nu\rho]\,\Lambda} \,, 
\end{equation}
where 
\begin{equation}
  \label{eq:def-G}
  \mathcal{G}_{\mu\nu\,\Lambda} = - \sqrt{|g|}\, 
  \varepsilon_{\mu\nu\rho\sigma}\,
  \frac{\partial \mathcal{L}}{\partial \mathcal{F}_{\rho\sigma}{}^\Lambda}
  \;.
\end{equation}
Here we use a metric with signature $(-,+,+,+)$ and
$\varepsilon_{0123}=1$. The discussion below is valid for any
space-time metric, and to simplify matters we will henceforth suppress
$g_{\mu\nu}$ and restrict ourselves to flat Minkowski space.
Obviously the set of equations (\ref{eq:eom-bianchi}) are invariant
under rotations of the $2n$-component array
$(\mathcal{F}^\Lambda,\mathcal{G}_\Lambda)$,
\begin{equation}
  \label{eq:em-duality}
  \pmatrix{\mathcal{F}^\Lambda\cr \mathcal{G}_\Lambda}\longrightarrow
  \pmatrix{U^\Lambda{}_\Sigma & Z^{\Lambda\Sigma} \cr 
   W_{\Lambda\Sigma} & V_\Lambda{}^\Sigma }   
  \pmatrix{\mathcal{F}^\Sigma\cr \mathcal{G}_\Sigma} \,.
\end{equation}
The new field strengths $\mathcal{G}_\Lambda$ can be written in the
form (\ref{eq:def-G}) with a new Lagrangian, provided that the matrix
in (\ref{eq:em-duality}) constitutes an element of the group
$\mathrm{Sp}(2n,\mathbb{R})$. Obviously these transformations are
generalizations of the duality transformations known from Maxwell
theory, which rotate the electric and magnetic fields and inductions
(for a review of electric/magnetic duality, see \cite{deWit:2001pz}).

We will employ an $\mathrm{Sp}(2n,\mathbb{R})$
covariant notation for the $2n$-dimensional symplectic indices
$M,N,\ldots$, such that $Z^M= (Z^\Lambda, Z_\Lambda)$. Likewise we use
vectors with lower indices according to $Y_M= (Y_\Lambda,Y^\Lambda)$,
transforming according to the conjugate representation so that
$Z^M\,Y_M$ is invariant. Our conventions are such that the
$\mathrm{Sp}(2n,\mathbb{R})$ invariant skew-symmetric tensor $\Omega_{MN}$
takes the form,
\begin{equation}
  \label{eq:omega}
  \Omega = \left( \begin{array}{cc} 
0 & {\bf 1}\\ \!-{\bf 1} & 0 
\end{array} \right)  \;.  
\end{equation}
The conjugate matrix $\Omega^{MN}$ is defined by
$\Omega^{MN}\Omega_{NP}= - \delta^M{}_P$. 

In the following we will be dealing with Lagrangians that are at most
quadratic in the field strengths, although our methods can also be
applied to more complicated Lagrangians. In addition the Lagrangian may
depend on other fields. Let us consider a generalization of the
kinetic term depending on a (possibly field-dependent) symmetric
tensor $\mathcal{N}_{\Lambda\Sigma}$,
\begin{eqnarray}
  \label{eq:quadratic-L}
\mathcal{L}_{0} &=&
-\ft14 \, \mathrm{i} \Big\{\mathcal{N}_{\Lambda\Sigma} 
\, \mathcal{F}_{\mu\nu}^+{}^{\Lambda}\, \mathcal{F}^{+\mu\nu\Sigma}  -
 \bar\mathcal{N}_{\Lambda\Sigma} \, \mathcal{F}_{\mu\nu}^-{}^{\Lambda}\, 
  \mathcal{F}^{-\mu\nu\Sigma} \Big\} \nonumber \\[1ex]
&=&{}
 \ft14 \, {\cal I}_{\Lambda\Sigma}\,\mathcal{F}_{\mu\nu}{}^{\Lambda} 
\mathcal{F}^{\mu\nu\,\Sigma} 
+\ft 18 {\cal R}_{\Lambda\Sigma}\;\varepsilon^{\mu\nu\rho\sigma} 
\mathcal{F}_{\mu\nu}{}^{\Lambda} 
\mathcal{F}_{\rho\sigma}{}^{\Sigma}  \;,  
\end{eqnarray}
where the $\mathcal{F}^\pm_{\mu\nu}$ are complex selfdual combinations
with eigenvalue $\mp\mathrm{i}$ normalized such that
$\mathcal{F}_{\mu\nu}= \mathcal{F}^+_{\mu\nu} +
\mathcal{F}^-_{\mu\nu}$; $\mathcal{R}$ and $\mathcal{I}$ denote the
real and imaginary parts of $\mathcal{N}$ and play the role of
generalized theta angles and coupling constants, respectively. 

Upon an electric/magnetic duality transformation (\ref{eq:em-duality})
one finds an alternative Lagrangian of the same form but with a
different expression for $\mathcal{N}_{\Lambda\Sigma}$,
\begin{equation}
  \label{eq:N-transform}
  \mathcal{N}_{\Lambda\Sigma}\longrightarrow 
  (V \mathcal{N} + W )_{\Lambda\Gamma} \,
  [(U + Z \mathcal{N})^{-1}]^\Gamma{}_\Sigma \,.
\end{equation}
This result follows from requiring consistency between
(\ref{eq:def-G}) and (\ref{eq:em-duality}).  The symmetry of the new
$\mathcal{N}$ is ensured by the fact that (\ref{eq:em-duality})
belongs to $\mathrm{Sp}(2n,\mathbb{R})$. Two Lagrangians with tensors
$\mathcal{N}_{\Lambda\Sigma}$ related via (\ref{eq:N-transform}), are
equivalent by electric/magnetic duality.  However, if the tensor
$\mathcal{N}_{\Lambda\Sigma}$ depends on fields whose transformations
induce precisely a change of $\mathcal{N}_{\Lambda\Sigma}$ of the form
(\ref{eq:N-transform}), then we may be dealing with an invariance of
the combined field equations and Bianchi
identities~\cite{Gaillard:1981rj}. Of course, this invariance is only
realized provided that also the field equations associated with fields
other than the gauge fields, will respect the invariance. For
instance, the corresponding transformations of the scalar fields
should leave the scalar kinetic term invariant. Since this term takes
the form of a non-linear sigma model, the transformations must
constitute isometries of the target space manifold.

The above transformations can be realized for more general
Lagrangians. In particular we can introduce a moment coupling of the
form, 
\begin{equation}
  \label{eq:moment-coupling}
  \mathcal{L}_{\mathrm{m}}= \mathcal{F}_{\mu\nu}^{+\Lambda}
  \,\mathcal{O}_\Lambda^{+\mu\nu} + \mathcal{F}_{\mu\nu}^{-\Lambda}
  \,\mathcal{O}_\Lambda^{-\mu\nu} \,,
\end{equation}
where $\mathcal{O}_{\mu\nu\,\Lambda}^\pm$ depends on matter fields and
is usually bilinear in spinor fields. Equivalent Lagrangians are now
defined in terms of tensors $\mathcal{N}$ related according to
(\ref{eq:N-transform}) and tensors $\mathcal{O}^{\pm}$ related according to 
\begin{equation}
  \label{eq:O-em-dual}
   \mathcal{O}^{+}_{\mu\nu\,\Lambda} \longrightarrow
   \mathcal{O}^{+}_{\mu\nu\,\Sigma} \, [(U +
   Z\mathcal{N})^{-1}]^\Sigma{}_\Lambda\,, 
\end{equation}
and likewise for $\mathcal{O}_{\mu\nu\,\Lambda}^-$, but with
$\mathcal{N}$ replaced by its complex conjugate. In order to have an
invariance, the transformations (\ref{eq:O-em-dual}) should be induced
by the transformations of the fields on which the tensors
$\mathcal{O}_{\mu\nu\,\Lambda}^\pm$  depend. In the presence of the
moment coupling (\ref{eq:moment-coupling}), it is advantageous to also
include the following matter term into the
Lagrangian~\cite{deWit:2001pz}, 
\begin{equation}
  \label{eq:O2-term}
  \mathcal{L}^\prime{}_{\mathrm{m}}=
  \ft12 [\mathcal{I}^{-1}]^{\Lambda\Sigma} \, \mathcal{O}_{\mu\nu\,\Lambda}
  \,\mathcal{O}^{\mu\nu}{}_\Sigma \,,
\end{equation}
with $\mathcal{O}_{\Lambda}=\mathcal{O}^{+}_{\Lambda}+
\mathcal{O}^{-}_{\Lambda}$. While this term is itself not invariant,
it ensures that the unspecified remaining terms in the total
Lagrangian will be separately invariant.

In this paper we will be dealing with the full group of invariances
which we denote by $\mathrm{G}$. According to the above, the invariance
transformations that act on the vector fields should always comprise a
subgroup of the electric/magnetic duality group.  This implies that a
$2n$-dimensional representation of $\mathrm{G}$ should exist with
generators $(t_\alpha)_M{}^N$, where the indices $\alpha$ label the
generators, satisfying
\begin{equation}
  \label{eq:symplectic}
  (t_\alpha)_{[M}{}^P\,\Omega_{N]P} =0\,.  
\end{equation}
On a $2n$-dimensional symplectic
vector $Z_M$, such a transformation takes the form $\delta Z_M=
\Lambda^\alpha \,(t_\alpha)_M{}^N\,Z_N$, where the matrix decomposes
according to
\begin{equation}
  \label{eq:sympl-infi}
  \Lambda^\alpha \,(t_\alpha)_M{}^N= \pmatrix{b_\Lambda{}^\Sigma &
  c_{\Lambda\Sigma} \cr  d^{\Lambda\Sigma} &
  -(b^{\mathrm{T}})^\Lambda{}_\Sigma} \,, 
\end{equation}
with $c_{\Lambda\Sigma}=c_{\Sigma\Lambda}$ and
$d^{\Lambda\Sigma}=d^{\Sigma\Lambda}$. The matrices $b,c,d$ comprise
at most $n(2n+1)$ independent parameters, which is consistent with the
fact that we are dealing with a subgroup of
$\mathrm{Sp}(2n,\mathbb{R})$. Observe that the above conventions are
such that the infinitesimal form of the matrix in
(\ref{eq:em-duality}) reads as follows, $U\approx\mathrm{1}-
b^{\mathrm{T}}$, $V\approx\mathrm{1} + b$, $W\approx - c$, $Z\approx
-d$.

For continuous invariances there is a simple way to determine the
explicit form of the submatrices $b$, $c$ and $d$. Namely one
sandwiches (\ref{eq:sympl-infi}) with the symplectic vectors
$(\mathcal{F}_{\mu\nu}{}^\Lambda,\mathcal{G}_{\mu\nu\,\Lambda})$ and
its dual $(\mathcal{G}_{\rho\sigma\,\Sigma},
- \mathcal{F}_{\rho\sigma}{}^\Sigma)$ and contracts over
$\varepsilon^{\mu\nu\rho\sigma}$. The resulting expression must vanish
(see eq.~12 of~\cite{deWit:2001pz}), which is a rather stringent
condition on the generators.

We caution the reader that at this point the invariance applies
to the combined equations of motion and the Bianchi identities, while the
Lagrangian is in general {\it not} invariant. In principle there is
nothing wrong with this and supergravity theories have provided many
examples of theories where this situation is realized.

The dual field strengths~(\ref{eq:def-G}) derived from the
combined Lagrangian~(\ref{eq:quadratic-L}), (\ref{eq:moment-coupling})
read
\begin{equation}
\label{eq:Gex}
  {\cal G}_{\mu\nu\,\Lambda} =
{\cal R}_{\Lambda\Gamma} {\cal F}_{\mu\nu}{}^{\Gamma}
-\ft12\varepsilon_{\mu\nu\rho\sigma}\,
{\cal I}_{\Lambda\Gamma}\, {\cal F}{}^{\rho\sigma\,\Gamma}
- \varepsilon_{\mu\nu\rho\sigma}\,{\cal O}^{\rho\sigma}{}_{\Lambda}\;.
\end{equation}
The relation (\ref{eq:Gex}) is consistent
with all the transformation rules given previously. So far we only
introduced electric gauge fields $A_\mu{}^\Lambda$, but at this stage
one can also introduce their magnetic duals $A_{\mu\,\Lambda}$
associated with these dual field strengths $\mathcal{G}_{\mu\nu
  \,\Lambda}$, by writing $\mathcal{G}_{\mu\nu \,\Lambda} \equiv 2
\,\partial_{[\mu} A_{\nu]\Lambda}$. The invariance group $\mathrm{G}$
mixes the two types of field strengths, as follows from
(\ref{eq:em-duality}).  Therefore the generators should be viewed as
generalized charges that contain both electric and magnetic
components. Switching on a gauge coupling constant may thus require
both electric and magnetic vector potentials.

\section{Gauging with electric and magnetic potentials}
We now introduce gauge couplings into the Lagrangian without
restricting ourselves to only electric charges. Hence we introduce
gauge fields $A_\mu{}^M$ which decompose into electric gauge fields
$A_\mu{}^\Lambda$ and magnetic gauge fields $A_{\mu\,\Lambda}$. Of
course, usually only a subset of these fields will be involved in the
gauging. Introducing magnetic gauge fields could lead to additional
propagating degrees of freedom. We will discuss in due course how to
avoid this. 

The gauge group must be embedded into the rigid invariance group. This
is done by means of an embedding tensor $\Theta_M{}^\alpha$ which
determines the decomposition of the gauge group generators $X_M$ into
the generators associated with the rigid invariance group $\mathrm{G}$,
\begin{equation}
  \label{eq:X-into-t}
  X_M = \Theta_M{}^\alpha \,t_\alpha \,.
\end{equation}
Not all the gauge fields have to be involved in the gauging, so
generically the embedding tensor projects out certain combinations of
gauge fields; the rank of the tensor determines the dimension of
the gauge group, up to central extensions. Decomposing the embedding tensor as
$\Theta_M{}^\alpha = (\Theta_\Lambda{}^\alpha,
\Theta^{\Lambda\,\alpha})$, covariant derivatives take the form,
\begin{equation}
  \label{eq:cov-der}
   D_\mu\equiv 
  \partial_{\mu}-g A_{\mu}{}^{M} X_M = \partial_{\mu}-g
A_{\mu}{}^{\Lambda}\Theta_{\Lambda}{}^{\alpha} \,t_{\alpha} -g
A_{\mu\,\Lambda}\Theta^{\Lambda\,\alpha} \,t_{\alpha} \;.  
\end{equation}
As stressed in section~1, the embedding tensor is treated as a
spurionic object, which can then be assigned to a (not necessarily
irreducible) representation of the rigid invariance group $\mathrm{G}$.

{}From our experience with supergravity, we know that a number of
($\mathrm{G}$-covariant) constraints must be imposed on the embedding
tensor. We introduce two such constraints quadratic in the embedding
tensor,
\begin{eqnarray}
  f_{\alpha\beta}{}^{\gamma}\, \Theta_{M}{}^{\alpha}\,\Theta_{N}{}^{\beta}
+(t_{\alpha})_{N}{}^{P}\,\Theta_{M}{}^{\alpha}\Theta_{P}{}^{\gamma} &=&0\,, 
  \label{eq:clos}  \\[1ex]
\Omega^{MN}\,\Theta_{M}{}^{\alpha}\Theta_{N}{}^{\beta}~=~0
\;\;\Longleftrightarrow\; \;
\Theta^{\Lambda\,[\alpha}\Theta_{\Lambda}{}^{\beta]} &=&0 \;,
\label{eq:quad}
\end{eqnarray}
where the $f_{\alpha\beta}{}^\gamma$ are the structure constants
associated with the group $\mathrm{G}$.  The first constraint is
required by the closure of the gauge group generators. Indeed, from
(\ref{eq:clos}) it follows that the gauge algebra generators close
according to
\begin{equation}
  \label{eq:closure}
  {}[X_{M},X_{N}] = -X_{MN}{}^{P}\,X_{P} \;,
\end{equation}
where the structure constants of the gauge group coincide with
$X_{MN}{}^{P}\equiv \Theta_{M}{}^{\alpha} \,(t_{\alpha})_{N}{}^{P}$ up
to terms that vanish upon contraction with the embedding tensor
$\Theta_P{}^\alpha$.  We recall that the $X_{MN}{}^P$ generate a
subgroup of $\mathrm{Sp}(2n,\mathbb{R})$ in the $(2n)$-dimensional
representation, so that $X_{M\Lambda}{}^\Sigma=
-X_{M}{}^\Sigma{}_\Lambda$, $X_{M \Lambda\Sigma}=X_{M\Sigma\Lambda}$
and $X_{M}{}^{\Lambda\Sigma}=X_{M}{}^{\Sigma\Lambda}$. Note that
(\ref{eq:clos}) also establishes the gauge invariance of the embedding
tensor. The second quadratic constraint (\ref{eq:quad}) implies that
the charges are mutually local, so that an electric/magnetic duality
exists that will convert all the charges to electric ones.  

In addition, we impose the following ($\mathrm{G}$-covariant) linear
constraint on $\Theta_{M}{}^{\alpha}$,
\begin{equation}
  \label{eq:lin}
  X_{(MN}{}^{Q}\,\Omega_{P)Q} =0 
\Longrightarrow  \left\{
\begin{array}{l}
X^{(\Lambda\Sigma\Gamma)}=0\,,\\[.2ex]
2X^{(\Gamma\Lambda)}{}_{\Sigma}= 
X_{\Sigma}{}^{\Lambda\Gamma}\,, \\[.2ex]
X_{(\Lambda\Sigma\Gamma)}=0\,,\\[.2ex]
2X_{(\Gamma\Lambda)}{}^{\Sigma}=
X^{\Sigma}{}_{\Lambda\Gamma}\,,
\end{array}
\right.
\end{equation}
which implies that we suppress a number of independent irreducible
representations that are generically contained in the embedding
tensor. 

Obviously one can impose additional constraints on the embedding
tensor, but the above set is probably the minimal one. The 
constraints (\ref{eq:clos}) and (\ref{eq:lin}) are known from maximal
$N=8$ supergravity 
\cite{deWit:2002vt,deWit:2003hr,dWST4}, where (\ref{eq:lin}) is
required by the supersymmetry of the action. These two constraints imply the
validity of the third one (\ref{eq:quad}). The relation between the
two quadratic constraints turns out to be a more generic feature, as we
can see by symmetrizing the constraint (\ref{eq:clos}) in
$(MN)$ and by using the linear constraint~(\ref{eq:lin})  and
(\ref{eq:symplectic}). This leads to 
\begin{eqnarray}
  \label{eq:constraint-eq}
\Omega^{MN}\,\Theta_{M}{}^{\alpha}\Theta_{N}{}^{\beta}\,
(t_{\beta})_{P}{}^{Q}&=&0 \;,  
\end{eqnarray}
which shows that for nonvanishing $(t_{\beta})_{P}{}^{Q}$ the second
quadratic constraint~(\ref{eq:quad}) is in fact a consequence of the
other constraints just as for the $N=8$ theory. Only for those
generators $t_{\alpha}$ that have a trivial action on the vector
fields, does (\ref{eq:quad}) represent an independent constraint.
This happens only when the symmetry group of the ungauged theory
factorizes into a product of several groups. We will encounter this
situation later in section~5.

As a further consequence of~(\ref{eq:lin}) one finds that
\begin{equation}
  \label{eq:Z-d}
  X_{(MN)}{}^{P}= Z^{P,\alpha}\, d_{\alpha \,MN} \;, 
\end{equation}
with
\begin{eqnarray}
  \label{eq:def-Z-d}
d_{\alpha\, MN} &\equiv& (t_\alpha)_M{}^P\, \Omega_{NP}\,,\nonumber\\
Z^{M,\alpha}&\equiv&\ft12\Omega^{MN}\Theta_{N}{}^{\alpha}
\quad
\Longrightarrow
\quad
\left\{
\begin{array}{rcr}
Z^{\Lambda,\alpha} &=& \ft12\Theta^{\Lambda\alpha} \,,\\[1ex]
Z_{\Lambda}{}^{,\alpha} &=& -\ft12\Theta_{\Lambda}{}^{\alpha} \,.
\end{array}
\right.
\end{eqnarray}
The tensor $d_{\alpha\,MN}$ defines a $\mathrm{G}$-invariant tensor
symmetric in $(MN)$. The gauge invariant tensor $Z^{M,\alpha}$ will
serve as a projector on the tensor fields to be introduced in the
following~\cite{deWit:2005hv}.  By virtue of the constraint
(\ref{eq:quad}), we have
\begin{equation}
  \label{eq:Z-Theta}
  Z^{M,\alpha} \,\Theta_M{}^\beta =0\,.
\end{equation}

Let us return to the closure relation (\ref{eq:closure}). Although
the left-hand side is antisymmetric in $M$ and $N$, this does not
imply that $X_{MN}{}^P$ is antisymmetric as well, but only that its symmetric
part vanishes upon contraction with the embedding tensor. Indeed, this is
reflected by (\ref{eq:Z-d}) and (\ref{eq:Z-Theta}). Consequently,
the Jacobi identity holds only modulo terms that vanish upon
contraction with the embedding tensor, as is shown by 
\begin{equation}
\label{Jacobi-X}
 {X_{[MN]}{}^P\, X_{[QP]}{}^R + X_{[QM]}{}^P\, X_{[NP]}{}^R  + X_{[NQ]}{}^P \,
 X_{[MP]}{}^R} = - Z^{R,\alpha}\,d_{\alpha\,P[Q}\, X_{MN]}{}^P \,.  
\end{equation}
To compensate for this lack of closure and, at the same time, to avoid
unwanted degrees of freedom, we introduce an extra gauge invariance
for the gauge fields, in addition to the usual nonabelian gauge
transformations, 
\begin{equation}
  \label{eq:A-var}
\delta A_\mu{}^M =  D_\mu\Lambda^M- g\,Z^{M,\alpha}\,\Xi_{\m\,\alpha}\,,  
\end{equation}
where the $\Lambda^M$ are the gauge transformation parameters and the
covariant derivative reads, $D_\mu\Lambda^M =\partial_\mu\Lambda^M +
g\, X_{PQ}{}^M\,A_\mu{}^P\Lambda^Q$. The transformations proportional
to $\Xi_{\m\,\alpha}$ enable one to gauge away those vector fields
that are in the sector of the gauge generators $X_{MN}{}^P$ where the
Jacobi identity is not satisfied (this sector is perpendicular to the
embedding tensor).\footnote{
  \label{ftn}
  Here we modified the gauge field transformation rules as compared to
  earlier publications \cite{deWit:2004nw,deWit:2005hv}, which simply
  amounts to a redefinition, $\Xi_{\mu\,\alpha}\to
  \Xi_{\mu\,\alpha}- d_{\alpha PQ} A_\mu{}^P\Lambda^Q$. This
  modification will lead to certain simplifications later on.}
Because the Jacobi identity is not satisfied and because of the extra
gauge transformations, the usual field strength, which follows from
the Ricci identity, $[D_\mu,D_\nu]= - g
\mathcal{F}_{\mu\nu}{}^M\,X_M$,
\begin{equation}
  \label{eq:field-strength}
  {\cal  F}_{\mu\nu}{}^M =\pa_\m A_\n{}^M -\pa_\n A_\m{}^M + g\,
X_{[NP]}{}^M \,A_\m{}^N A_\n{}^P \,,
\end{equation}
is not fully covariant.\footnote{
   Observe that the covariant derivative is invariant under the tensor
   gauge transformations, so that the field strengths contracted with
   $X_M$ are in fact covariant. } 
Therefore we define a modified field strength, 
\begin{equation}
  \label{eq:modified-fs}
{\cal H}_{\m\n}{}^M= 
{\cal F}_{\mu\nu}{}^M  
+ g\, Z^{M,\alpha} \,B_{\m\n\,\alpha}\;,
\end{equation}
where we introduce the tensor fields $B_{\m\n\,\alpha}$, whose transformation
rules will be defined such that the ${\cal H}_{\m\n}{}^M$ transform
covariantly, 
\begin{equation}
  \label{eq:delta-H}
\delta\mathcal{H}_{\mu\nu}{}^M = -
g\,X_{PN}{}^M\,\Lambda^P\mathcal{H}_{\mu\nu}{}^N\,.   
\end{equation}
This leads to the following result for the transformation rule of
$B_{\m\n\,\alpha}$,\footnote{
   This result is taken from \cite{deWit:2005hv}, but it takes a
   different form due to the redefinition of $\Xi_{\mu \,\alpha}$,
   discussed in footnote~\ref{ftn}.} 
\begin{equation}
  \label{eq:B-transf-0}
   \delta B_{\mu\nu\,\alpha} = 2\,
   D_{[\mu} \Xi_{\nu]\alpha} + 2\,d_{\alpha\,MN}A_{[\mu}{}^M \delta
   A_{\nu]}{}^N  - 2\,d_{\alpha\, MN} \Lambda^M
   \mathcal{H}_{\mu\nu}{}^N 
   \,,
\end{equation}
up to terms that vanish under contraction with $Z^{M,\alpha}$. As it
turns out, we do not need these contributions as variations of the
tensor field in the final Lagrangian will always be multiplied by
$Z^{\Lambda,\alpha}$. 
 The
relevant variation is therefore,
\begin{eqnarray}
  \label{eq:B-transf-1}
   \Theta^{\Lambda\alpha} \delta B_{\mu\nu\,\alpha} &=&
   2\,\Theta^{\Lambda\alpha} \Big[
   D_{[\mu} \Xi_{\nu]\alpha} + d_{\alpha\,
   MN}A_{[\mu}{}^M \delta A_{\nu]}{}^N\Big] \nonumber \\
  &&{}
    - 2\,\Lambda^M\Big[
   X^\Lambda{}_{M\Sigma}\, \mathcal{H}_{\mu\nu}{}^\Sigma -
   X^\Lambda{}_M{}^\Sigma \,\mathcal{H}_{\mu\nu\Sigma}\Big] \;, 
\end{eqnarray}
where we made use of (\ref{eq:def-Z-d}). 

In passing we note that the covariant field strength of the tensor
fields is known and given by \cite{deWit:2005hv},
\begin{equation}
  \label{eq:tensor-H}
{\cal H}^{{(3)}}_{\mu\nu\rho\,\alpha} \equiv 3\, D_{[\mu}
B_{\nu\rho]\,\alpha} +6 \,d_{\alpha\, MN}\,A_{[\mu}{}^{M}
(\partial_{\nu} A_{\rho]}{}^N+ \ft13 g X_{[RS]}{}^{N}
A_{\nu}{}^{R}A_{\rho]}{}^{S}) \;,  
\end{equation}
up to terms that vanish when contracted with $Z^{M,\alpha}$. 
The vector and tensor field strengths satisfy the generalized Bianchi
identities (the tensor identity holds upon contraction with
$Z^{M,\alpha}$),
\begin{eqnarray}
  \label{eq:vt-bianchi}
  Z^{M,\alpha} D^{\vphantom{l}}_{[\mu}{\cal
  H}^{{(3)}}_{\nu\rho\sigma]\alpha} &=& 3g\,X_{PQ}{}^{M}\, {\cal
  H}_{[\mu\nu}{}^{P}\,{\cal H}_{\rho\sigma]}{}^{Q} \;,
\\[1ex]
D_{[\mu}{\cal H}_{\nu\rho]}{}^{M} &=& \ft13g\,Z^{M,\alpha}\,{\cal
  H}^{{(3)}}_{\mu\nu\rho\,\alpha} \;, 
  \label{eq:vt-bianchi2}
\end{eqnarray}
with the covariant derivatives $D{\cal H}{}^{M} =\partial{\cal H}^{M}
+g\, X_{PQ}{}^{M}A^{P}\,{\cal H}{}^{Q}$ and $D{\cal H}_{\alpha} =
\partial{\cal H}_{\alpha} +g \,\Theta_{M}{}^{\gamma}
f_{\gamma\alpha}{}^{\beta}\,A^{M} {\cal H}_{\beta}$.

The next steps are rather obvious. Namely one covariantizes the combined
Lagrangian (\ref{eq:quadratic-L}), (\ref{eq:moment-coupling}) and adds
a topological term that involves the tensor and vector fields.
However, in four space-time dimensions this will not directly lead to
the correct solution and further modifications will be required. This
is related to the fact that the rigid invariance $\mathrm{G}$ was not
necessarily an invariance of the initial Lagrangian, but of the
combined equations of motion and the Bianchi identities. In this
section we will therefore carry out these first steps and exhibit the
problematic features of this intermediate result. In addition we will
show that the purely electric gaugings do not suffer from any of these
problems. In section~4 we will then introduce the complete gauge
invariant Lagrangian and transformation rules. In all of this the
embedding tensor constraints play a crucial role. 

Covariantizing the combined Lagrangian (\ref{eq:quadratic-L}),
(\ref{eq:moment-coupling}) leads to 
\begin{equation}
  \label{eq:L-cov}
  \mathcal{L}_{0}+\mathcal{L}_{\rm m} ~=~
 \ft14 \, {\cal I}_{\Lambda\Sigma}\,\mathcal{H}_{\mu\nu}{}^{\Lambda} 
\mathcal{H}^{\mu\nu\,\Sigma} 
+\ft 18 {\cal R}_{\Lambda\Sigma}\;\varepsilon^{\mu\nu\rho\sigma} 
\mathcal{H}_{\mu\nu}{}^{\Lambda} 
\mathcal{H}_{\rho\sigma}{}^{\Sigma}  
+
\mathcal{H}_{\mu\nu}{}^{\Lambda}
  \,\mathcal{O}^{\mu\nu}{}_\Lambda\;.
\end{equation}
However, this Lagrangian is not invariant for the same reason as the
original Lagrangian was not invariant. To see this one makes use of the
infinitesimal gauge transformations, which for
$\mathcal{N}_{\Lambda\Sigma}$ and $\mathcal{O}^+_{\mu\nu\,\Lambda}$
take the form, 
\begin{eqnarray}
  \label{eq:gauge-var-N-O}
  \delta\mathcal{N}_{\Lambda\Sigma} &=& g\,\Lambda^M\Big[
  - X_{M\Lambda\Sigma} +2\, 
  X_{M(\Lambda}{}^\Gamma \mathcal{N}_{\Sigma) \Gamma} +
  \mathcal{N}_{\Lambda\Gamma} \,X_M{}^{\Gamma\Xi}\,
  \mathcal{N}_{\Xi\Sigma} \Big]  \;, 
  \nonumber\\
  \delta O^+_{\mu\nu\,\Lambda}&=& g\, \Lambda^M\,
  O^+_{\mu\nu\,\Sigma}\Big[X_{M\Lambda}{}^\Sigma  + 
  X_{M}{}^{\Sigma\Gamma} \,\mathcal{N}_{\Gamma\Lambda}\Big] \;. 
\end{eqnarray}
Using the variations (\ref{eq:delta-H}) and (\ref{eq:gauge-var-N-O}),
one derives
\begin{eqnarray}
  \label{eq:delta-cov-L}
  \delta(\mathcal{L}_{0}+\mathcal{L}_{\rm m}+\mathcal{L}^\prime{}_{\rm
  m})  &=&{} -\ft18 g\,\Lambda^M
  X_{M\Lambda\Sigma}
  \,\mathcal{H}_{\mu\nu}{}^\Lambda  \mathcal{H}_{\rho\sigma}{}^\Sigma
  \,\varepsilon^{\mu\nu\rho\sigma} 
  \nonumber \\ 
  &&{} 
  + \ft18 g\,\Lambda^M
  X_M{}^{\Lambda\Sigma} \,\mathcal{G}_{\mu\nu\Lambda}
  \,\mathcal{G}_{\rho\sigma\Sigma}   \,\varepsilon^{\mu\nu\rho\sigma} 
  \nonumber \\
  &&{}
  - \ft14 g\,\Lambda^M
  X_M{}^{\Lambda\Sigma}  \,\mathcal{G}_{\mu\nu\Lambda}
  \,\mathcal{H}_{\rho\sigma\Sigma}\,\varepsilon^{\mu\nu\rho\sigma} \;, 
\end{eqnarray}
where 
\begin{equation}
  \label{eq:def-G(H)}
  {\mathcal G}_{\mu\nu\,\Lambda} = {\cal R}_{\Lambda\Gamma} {\cal
  H}_{\mu\nu}{}^{\Gamma} -\ft12\varepsilon_{\mu\nu\rho\lambda}\, 
  {\cal I}_{\Lambda\Gamma}\, {\cal H}{}^{\rho\lambda\,\Gamma} -
  \varepsilon_{\mu\nu\rho\lambda}\,
  {\cal O}^{\rho\lambda}{}_{\Lambda}\;,   
\end{equation}
is a covariant version of (\ref{eq:Gex}).  Furthermore we note that
the variation with respect to $B_{\mu\nu\alpha}$ leads also to the
tensor ${\mathcal G}_{\mu\nu\,\Lambda}$,
\begin{equation}
  \label{eq:delta-B-cov-L}
  \delta(\mathcal{L}_{0}+\mathcal{L}_{\rm m}+\mathcal{L}^\prime{}_{\rm
  m})  = \ft18 g\,\varepsilon^{\mu\nu\rho\sigma} 
  \,\mathcal{G}_{\mu\nu\Lambda} \,\Theta^{\Lambda\alpha} \,\delta
  B_{\mu\nu\,\alpha} \,. 
\end{equation}

On the basis of the results found for maximal supergravity in five
space-time dimensions \cite{deWit:2004nw} and the more general
considerations presented in \cite{deWit:2005hv}, we introduce a
topological term,
$g\,\Theta^{\Lambda\,\alpha}\,\varepsilon^{\mu\nu\rho\sigma}
B_{\mu\nu\,\alpha}\, \partial_\rho A_{\sigma\,\Lambda}$, which only
contains the magnetic potentials so that the tensor field will
decouple for purely electric gaugings (the need for this will be
explained below). The term $\partial_{[\rho} A_{\sigma]\Lambda}$
constitutes the first term of the field strength
$\mathcal{H}_{\rho\sigma\Lambda}$, which suggests that its completion
will at least involve the terms,
\begin{equation}
  \label{eq:LB}
  {\cal L}_{{\rm top},B}= -\ft18 g 
  \varepsilon^{\mu\nu\rho\sigma}\,
  \Theta^{\Lambda\alpha}\,B_{\mu\nu\,\alpha} \,
  \Big(2\,\partial_{\rho} A_{\sigma\,\Lambda} + g
   X_{MN\,\Lambda} \,A_\rho{}^M A_\sigma{}^N
  -\ft14g\, \Theta_{\Lambda}{}^{\beta}B_{\rho\sigma\,\beta} \Big)\;,
\end{equation}
so that its variation with
respect to $\delta B_{\mu\nu\alpha}$ is just proportional to 
$\mathcal{H}_{\mu\nu\Lambda}$, 
\begin{equation}
  \label{eq:delta-LtopB}
  \delta\mathcal{L}_{{\rm top},B} = -\ft18 g \, 
  \varepsilon^{\mu\nu\rho\sigma}\,\mathcal{H}_{\mu\nu\Lambda} 
  \, \Theta^{\Lambda\alpha}\,\delta B_{\rho\sigma\,\alpha} \, \,. 
\end{equation}
Note that the tensor
$\Theta^{\Lambda\alpha}\,\Theta_{\Lambda}{}^{\beta}$ that multiplies
the term quadratic in $B_{\mu\nu\Lambda}$ is symmetric in
$(\alpha,\beta)$, by virtue of the constraint (\ref{eq:quad}).
General variations of (\ref{eq:LB}) can be written as follows,
\begin{equation}
  \label{eq:delta-Ltop}
  \delta\mathcal{L}_{{\rm top},B} = -\ft18 g \, 
  \varepsilon^{\mu\nu\rho\sigma}\, \Big[
 \mathcal{F}_{\mu\nu\Lambda} \,\Theta^{\Lambda\alpha} \,\delta
 B_{\rho\sigma\,\alpha} + \delta\mathcal{H}_{\mu\nu\Lambda} \, 
  \, \Theta^{\Lambda\alpha}\,B_{\rho\sigma\,\alpha} \Big]\, \,. 
\end{equation}
Substituting the various variations, one finds,
\begin{eqnarray}
  \label{eq:delta-top}
    \delta\mathcal{L}_{{\rm top},B} &=& {} \ft18 g\,\Lambda^M
  X_{M\Lambda\Sigma}\Big[\mathcal{H}_{\mu\nu}{}^\Lambda
  \mathcal{H}_{\rho\sigma}{}^\Sigma -\mathcal{F}_{\mu\nu}{}^\Lambda
  \mathcal{F}_{\rho\sigma}{}^\Sigma \Big] 
  \,\varepsilon^{\mu\nu\rho\sigma} 
  \nonumber \\ 
  &&{} 
  + \ft18 g\,\Lambda^M  X_M{}^{\Lambda\Sigma}\Big[
   \mathcal{H}_{\mu\nu}{}_\Lambda 
  \mathcal{H}_{\rho\sigma}{}_\Sigma  -\ft12 \mathcal{F}_{\mu\nu}{}_\Lambda 
  \mathcal{F}_{\rho\sigma}{}_\Sigma \Big] 
   \varepsilon^{\mu\nu\rho\sigma} 
  \nonumber \\[.5ex]
  &&{}
  + \ft18 g\,\Lambda^M
  X^\Lambda{}_{M\Sigma}  \,\mathcal{F}_{\mu\nu\Lambda}
  \,\mathcal{F}_{\rho\sigma}{}^\Sigma \,\varepsilon^{\mu\nu\rho\sigma}
  \;,  
\end{eqnarray}
up to terms of order $g^2$ that contain noncovariant terms
which depend explicitly on $A_\mu{}^M$. The constraints on the
embedding tensor are crucial for deriving the above results.  

Clearly at this stage the combined Lagrangian is not invariant as the
tensors $\mathcal{G}_{\mu\nu\Lambda}$ and
$\mathcal{H}_{\mu\nu\Lambda}$ are unrelated, although we note that the
terms quadratic in $\mathcal{H}_{\mu\nu}{}^M$ cancel when
$\mathcal{G}_{\mu\nu\Lambda}$ and $\mathcal{H}_{\mu\nu\Lambda}$ are
identified. This observation will be relevant later on. However, one
is then still left with the terms quadratic 
in $\mathcal{F}_{\mu\nu}{}^M$. We will exhibit in the next section how
these variations are cancelled. To pave the way and to verify the
consistency of the construction up to this point, let us briefly
consider purely electric gaugings, to appreciate the role of the
constraints and to establish that our formalism will remain in the
more conventional setting. For electric gaugings,
$\Theta^{\Lambda\alpha}=0$, so that the generators $X^{\Lambda}=0$. In
that case the constraint~(\ref{eq:lin}) reduces to
\begin{equation}
  \label{eq:conel}
  X_{\Sigma}{}^{\Lambda\Gamma}=0\;,
  \qquad
  X_{(\Lambda\Sigma)}{}^{\Gamma}=0\;,
  \qquad
  X_{(\Lambda\Sigma\Gamma)}=0
  \;.
\end{equation}
The remaining gauge generators, $X_{[\Lambda\Sigma]}{}^\Gamma=
X_\Lambda{}^\Gamma{}_\Sigma$ and $X_{\Lambda\Sigma\Gamma}$ satisfy the
Jacobi identity, because the right-hand side of (\ref{Jacobi-X})
vanishes. Hence the generators have a block-triangular form.  The
modified field strength~(\ref{eq:modified-fs}) for the electric vector
fields reduces to an ordinary nonabelian field strength 
\begin{equation}
  \label{eq:electric-case}
 {\cal H}_{\mu\nu}{}^\Lambda= 2\,\partial_{[\mu} A_{\nu]}{}^\Lambda +
g\,X_{\Sigma\Gamma}{}^\Lambda \,A_{\mu}{}^\Sigma A_{\nu}{}^\Gamma \;,
\end{equation}
which contains neither magnetic vector fields nor tensor fields.
The variation of the Lagrangian follows from (\ref{eq:delta-cov-L}),
where only the first term contributes. 
It was observed long ago in~\cite{deWit:1984px} that this variation 
can be cancelled by introducing the following Chern-Simons-like term
to the action,
\begin{equation}
\label{eq:top-el}  
  {\cal L}_{\rm top, electric} = -\ft13g\,
\varepsilon^{\mu\nu\rho\sigma}X_{\Omega\Xi\Sigma}\,
A_{\mu}{}^{\Omega}A_{\nu}{}^{\Xi} \Big(\partial_{\rho}
A_{\sigma}{}^{\Sigma}+ \ft38 g\, X_{\Lambda\Gamma}{}^{\Sigma}\,
A_{\rho}{}^{\Lambda} A_{\sigma}{}^{\Gamma} \Big) \;, 
\end{equation}
provided $X_{(\Lambda\Sigma\Gamma)}=0$, which is precisely the last
equation of~(\ref{eq:conel}).  This extends possible gauge
transformations to those with triangular embedding into the symplectic
group~(\ref{eq:em-duality}), the so-called Peccei-Quinn
transformations.

\section{Tensor fields and the topological term}
In this section we demonstrate that a general gauge invariant
Lagrangian exists with both electric and magnetic vector potentials.
This Lagrangian is an extension of the Lagrangian discussed
in the previous section, by 
Chern-Simons-like terms such as (\ref{eq:top-el}). The only
restriction will be that the embedding tensor $\Theta_{M}{}^{\alpha}$
is subject to the constraints (\ref{eq:clos}), (\ref{eq:quad}) and
(\ref{eq:lin}).

The first observation is that the variations of the total Lagrangian
bilinear in $\mathcal{H}_{\rho\sigma}{}_\Sigma$ and
$\mathcal{G}_{\mu\nu}{}_\Lambda$ combine into $X_M{}^{\Lambda\Sigma}
(\mathcal{H}_{\mu\nu\Lambda} - \mathcal{G}_{\mu\nu\Lambda})
(\mathcal{H}_{\rho\sigma\Sigma} - \mathcal{G}_{\rho\sigma\Sigma})$,
which, according to (\ref{eq:delta-B-cov-L}) and (\ref{eq:delta-LtopB}), 
can be removed by assigning an additional variation to $\delta
B_{\mu\nu\alpha}$ proportional to $(\mathcal{H}_{\mu\nu\Lambda} -
\mathcal{G}_{\mu\nu\Lambda})$. With this variation the modified
transformation rule for $B_{\mu\nu\alpha}$ reads,
\begin{eqnarray}
  \label{eq:B-transf-2}
   \Theta^{\Lambda\alpha} \delta B_{\mu\nu\,\alpha} &=&
   2\,\Theta^{\Lambda\alpha} \Big[
   D_{[\mu} \Xi_{\nu]\alpha} + d_{\alpha\,
   MN}A_{[\mu}{}^M \delta A_{\nu]}{}^N\Big] \nonumber \\
  &&{}
    - 2\,\Lambda^M\Big[
   X^\Lambda{}_{M\Sigma}\, \mathcal{H}_{\mu\nu}{}^\Sigma -
   X^\Lambda{}_M{}^\Sigma \,\mathcal{G}_{\mu\nu\Sigma}\Big] \;.
\end{eqnarray}
Apart from this modification the remaining transformation rules are
left unchanged, but one should be aware that (\ref{eq:delta-H}) receives
corrections induced by the modifications in (\ref{eq:B-transf-2}).
With these transformation rules, the variation of the total Lagrangian
takes the form
\begin{eqnarray}
\label{varL}
  \delta(\mathcal{L}_{0}+\mathcal{L}_{\rm m}+\mathcal{L}^\prime{}_{\rm
  m}+\mathcal{L}_{{\rm top},B})  &=&{}
   -\ft18 g\,\Lambda^M
  X_{M\Lambda\Sigma}\, \mathcal{F}_{\mu\nu}{}^\Lambda
  \mathcal{F}_{\rho\sigma}{}^\Sigma \,\varepsilon^{\mu\nu\rho\sigma}
  \nonumber \\ 
  &&{} 
    -\ft1{16} g\,\Lambda^M
  X_{M}{}^{\Lambda\Sigma}\, \mathcal{F}_{\mu\nu\,\Lambda}
  \mathcal{F}_{\rho\sigma\,\Sigma} \,\varepsilon^{\mu\nu\rho\sigma}
    \nonumber \\ 
  &&{} 
    +\ft18 g\,\Lambda^M
  X^{\Lambda}{}_{M\Sigma}\,
  \mathcal{F}_{\mu\nu\,\Lambda}\mathcal{F}_{\rho\sigma}{}^\Sigma
  \,\varepsilon^{\mu\nu\rho\sigma} \nonumber \\
  &&{} +\cdots  \;,  
\end{eqnarray}
where the dots denote noncovariant terms of order $g^{2}$
which depend explicitly on $A_\mu{}^M$. 
In order to finally cancel this variation, the topological
Lagrangian~(\ref{eq:LB}) must be extended to
\begin{eqnarray}
  \label{eq:Ltop}   
{\cal L}_{\rm top} &=&
-\ft18g\, \varepsilon^{\mu\nu\rho\sigma}\,
\Theta^{\Lambda\alpha}\,B_{\mu\nu\,\alpha} \,
\Big(2\,\partial_{\rho} A_{\sigma\,\Lambda} + g
X_{MN\,\Lambda} \,A_\rho{}^M A_\sigma{}^N
-\ft14g\Theta_{\Lambda}{}^{\beta}B_{\rho\sigma\,\beta}
\Big)
\nonumber\\[.9ex]
&&{}
-\ft13g\, \varepsilon^{\mu\nu\rho\sigma} X_{MN\,\Lambda}\,
A_{\mu}{}^{M} A_{\nu}{}^{N}
\Big(\partial_{\rho} A_{\sigma}{}^{\Lambda}
+\ft14 gX_{PQ}{}^{\Lambda} A_{\rho}{}^{P}A_{\sigma}{}^{Q}\Big)
\nonumber\\[.9ex]
&&{}
-\ft16g\, \varepsilon^{\mu\nu\rho\sigma}X_{MN}{}^{\Lambda}\,
A_{\mu}{}^{M} A_{\nu}{}^{N}
\Big(\partial_{\rho} A_{\sigma}{}_{\Lambda}
+\ft14 gX_{PQ\Lambda} A_{\rho}{}^{P}A_{\sigma}{}^{Q}\Big)
\;.
\end{eqnarray}
Straightforward but tedious computation then shows that
the variation of the extra terms 
precisely cancel the contributions~(\ref{varL}) such that
the sum
\begin{equation}
  \label{eq:Lvecten}
{\cal L}_{\rm VT} =
{\cal L}_{\rm 0}+{\cal L}_{\rm m}+
\mathcal{L}^\prime{}_{\rm  m}+{\cal L}_{\rm top}
\;,  
\end{equation}
is invariant under both vector and tensor gauge transformations
up to total derivatives. The constraints~(\ref{eq:clos}),
(\ref{eq:quad}), and (\ref{eq:lin}) are crucial in the derivation of
this result. The topological term~(\ref{eq:Ltop}) contains the
first-order term (\ref{eq:LB}) for the magnetic vector fields
$A_{\Lambda}$ and the tensor fields $B_{\alpha}$ and the Chern-Simons-like
term (\ref{eq:top-el}). Indeed, for an electric
gauging~($\Theta^{\Lambda\alpha}=0$) the Chern-Simons-like terms in
(\ref{eq:Ltop}) reduce to (\ref{eq:top-el}), while for a purely 
magnetic gauging ($\Theta_{\Lambda}{}^{\alpha}=0$), they take the form
\begin{eqnarray}
  \label{eq:top-magnetic}
{\cal L}_{\rm top,\,\,magnetic} &=&
-\ft18g\,\varepsilon^{\mu\nu\rho\sigma}\, \Theta^{\Lambda\alpha}\,
B_{\mu\nu\,\alpha}
\Big(2\,\partial_{\rho} A_{\sigma{}}{}_\Lambda 
+ g\,X^{\Sigma}{}_{M\Lambda} \,A_{\rho}{}_\Sigma A_{\sigma}{}^M\Big)
  \nonumber
\\[.5ex]
&&{}
-\ft1{6}g\,\varepsilon^{\mu\nu\rho\sigma}X^{\Omega\Xi\Sigma}\,
A_{\mu\,\Omega} A_{\nu\,\Xi}
\Big( \partial_{\rho} A_{\sigma\Sigma}+ \ft38g\,
X^{\Lambda\Gamma}{}_{\Sigma}\,
A_{\rho}{}_{\Lambda} A_{\sigma}{}_{\Gamma}
\Big)
\nonumber\\[.5ex]
&&{}
  -\ft14g\,\varepsilon^{\mu\nu\rho\sigma}X^{\Omega\Xi}{}_{\Sigma}\,
  A_{\mu\,\Omega} A_{\nu\,\Xi} 
   \Big( \partial_{\rho} A_{\sigma}{}^{\Sigma}
   +\ft14g\,X^{\Lambda\Gamma\Sigma} \,A_{\rho}{}_{\Lambda}
   A_{\sigma}{}_{\Gamma} \Big) \;. 
\end{eqnarray}

Summarizing, we have shown that the total Lagrangian, 
\begin{eqnarray}
\label{eq:Ltotal}
  {\cal L}_{{\rm VT}} &=&
\ft14 \, {\cal I}_{\Lambda\Sigma}\,\mathcal{H}_{\mu\nu}{}^{\Lambda} 
\mathcal{H}^{\mu\nu\,\Sigma} 
+\ft 18 {\cal R}_{\Lambda\Sigma}\;\varepsilon^{\mu\nu\rho\sigma} 
\mathcal{H}_{\mu\nu}{}^{\Lambda} 
\mathcal{H}_{\rho\sigma}{}^{\Sigma}  
  +\mathcal{H}_{\mu\nu}{}^{\Lambda}
  \mathcal{O}^{\mu\nu}{}\!_\Lambda
  +  \ft12 [\mathcal{I}^{-1}]^{\Lambda\Sigma} \, \mathcal{O}_{\mu\nu\,\Lambda}
  \mathcal{O}^{\mu\nu}{}\!_\Sigma
  \nonumber\\[.9ex]
&&{}
-\ft18g\, \varepsilon^{\mu\nu\rho\sigma}\,
\Theta^{\Lambda\alpha}\,B_{\mu\nu\,\alpha} \,
\Big(
2\,\partial_{\rho} A_{\sigma\,\Lambda} + g
X_{MN\,\Lambda} \,A_\rho{}^M A_\sigma{}^N
-\ft14g\Theta_{\Lambda}{}^{\beta}B_{\rho\sigma\,\beta} \Big)
\nonumber\\[.9ex]
&&{}
-\ft13g\, \varepsilon^{\mu\nu\rho\sigma}X_{MN\,\Lambda}\,
A_{\mu}{}^{M} A_{\nu}{}^{N}
\Big(\partial_{\rho} A_{\sigma}{}^{\Lambda}
+\ft14 gX_{PQ}{}^{\Lambda} A_{\rho}{}^{P}A_{\sigma}{}^{Q}\Big)
\nonumber\\[.9ex]
&&{}
-\ft16g\, \varepsilon^{\mu\nu\rho\sigma}X_{MN}{}^{\Lambda}\,
A_{\mu}{}^{M} A_{\nu}{}^{N}
\Big(\partial_{\rho} A_{\sigma}{}_{\Lambda}
+\ft14 gX_{PQ\Lambda} A_{\rho}{}^{P}A_{\sigma}{}^{Q}\Big)
\;,
\end{eqnarray}
is invariant under the vector and tensor gauge transformations
(\ref{eq:A-var}), (\ref{eq:gauge-var-N-O}) and (\ref{eq:B-transf-2}). 
It provides a unified description of electric and magnetic vector
fields as well as of tensor fields 
which  encompasses all possible gaugings. The gauge group is
characterized by the embedding tensor $\Theta_{M}{}^{\alpha}$ subject
to the constraints~(\ref{eq:clos}), (\ref{eq:quad}), and
(\ref{eq:lin}). Apart from these constraints the embedding of the
gauge group into the symplectic 
group~(\ref{eq:em-duality}) is arbitrary. Due to the 
presence of both electric and magnetic vector fields the gauge group
is no longer restricted to diagonal or triangular embeddings. 
The gaugings are thus not necessarily restricted to subgroups of the 
invariance group of the initial ungauged Lagrangian
but may include additional invariances of the combined
set of Bianchi identities and field equations.
Upon partial gauge fixing and integrating out fields one 
recovers the vector/tensor couplings previously presented in the
literature as particular examples of (\ref{eq:Ltotal}). 
We will illustrate this with a few examples in the next section.

The vector/tensor Lagrangian (\ref{eq:Ltotal}) can be amended by
additional matter couplings of the vector fields to scalar and fermion
fields 
\begin{equation}
  \label{eq:Lvtm}
  {\cal L} = {\cal L}_{{\rm VT}} +{\cal L}_{\rm matter}\;.
\end{equation}
In these matter couplings the electric and magnetic vector fields
enter exclusively via the covariant derivatives (\ref{eq:cov-der}) and
therefore take a symplectically covariant form. It is important to
note that due to~(\ref{eq:Z-Theta}) the covariant derivatives are
invariant under tensor gauge transformations.

Under the variations $A_{\mu}{}^{M} \rightarrow A_{\mu}{}^{M}+ \delta
A_{\mu}{}^{M}$, and  
$B_{\mu\nu\,\alpha}\rightarrow B_{\mu\nu\,\alpha}+\delta B_{\mu\nu\,\alpha}$,
the vector/tensor Lagrangian (\ref{eq:Ltotal}) changes as
\begin{eqnarray}
  \label{eq:dLvtm}
\lefteqn{\delta{\cal L}_{\rm VT} ~=}
\nonumber\\[1ex]
\quad&=&
-\ft18g\varepsilon^{\mu\nu\rho\sigma}\,
\Big(\Theta^{\Lambda\alpha}\,\delta B_{\mu\nu\,\alpha}
-2\,X^\Lambda{}_{M}{}^{\Sigma} A_{\mu}{}^{M} \delta A_{\nu\,\Sigma}
-2\, X^\Lambda{}_{M\Sigma} A_{\mu}{}^{M} \delta A_{\nu}{}^{\Sigma}\Big)
({\cal H} -\mathcal{G})_{\rho\sigma\,\Lambda} 
\nonumber\\[.5ex]
&&{}
-
\ft1{12}g\varepsilon^{\mu\nu\rho\sigma} \,\delta A_{\mu\,\Lambda}
\Big(\Theta^{\Lambda\alpha}\, {\cal H}^{(3)}_{\nu\rho\sigma\,\alpha} 
+6\,X^\Lambda{}_{M}{}^{\Sigma} A_{\nu}{}^{M} 
  ({\cal H} - \mathcal{G})_{\rho\sigma\,\Sigma}   
\Big)
\nonumber\\[.5ex]
&&{}
+\ft12\varepsilon^{\mu\nu\rho\sigma}
\delta A_{\mu}{}^{\Lambda}
\Big(\partial_{\nu} \mathcal{G}_{\rho\sigma\,\Lambda}
-gX_{M\Lambda}{}^{\Sigma}\, A_{\nu}{}^{M} \mathcal{G}_{\rho\sigma\Sigma}
+gX_{M \Sigma \Lambda}\, A_{\nu}{}^{M}  {\cal H}_{\rho\sigma}{}^{\Sigma}
\Big)
\;,
\end{eqnarray}
with $\mathcal{G}_{\mu\nu\,\Lambda}$ defined in~(\ref{eq:def-G(H)}).
{}From these variations one reads off the equations of motion 
resulting from  (\ref{eq:Lvtm}), 
\begin{eqnarray}
g\,\Theta^{\Lambda\alpha}\,
({\cal H}-\mathcal{G})_{\mu\nu\,\Lambda}
&=& 0\;,
\label{eq:eom1}
\\[1ex]
\ft1{12} g\, \varepsilon^{\mu\nu\rho\sigma}\,
\Big({\Theta^{\Lambda\alpha}\,\cal H}^{(3)}_{\nu\rho\sigma\,\alpha} 
+6\,X^\Lambda{}_{M}{}^{\Sigma}
A_{\nu}{}^{M} ({\cal H} - \mathcal{G})_{\rho\sigma\,\Sigma} 
\Big)
&=& g\, j^{\mu\,\Lambda} \;,
\label{eq:eom2}
\\[1ex]
{}-\ft12 \varepsilon^{\mu\nu\rho\sigma}
\Big(\partial_{\nu} \mathcal{G}_{\rho\sigma\,\Lambda}
-gX_{M\Lambda}{}^{\Sigma}\, A_{\nu}{}^{M} \mathcal{G}_{\rho\sigma\Sigma}
+gX_{M \Sigma}{}_{\Lambda}\, A_{\nu}{}^{M}  {\cal H}_{\rho\sigma}{}^{\Sigma}
\Big)
&=& g\, j^\mu{}_\Lambda \;,
\label{eq:eom3}
\end{eqnarray}
where $(j^{\mu\Lambda}, j^\mu{}_\Lambda)$ denote the magnetic and
electric current densities associated with ${\cal L}_{\rm matter}$,
which are defined by
\begin{equation}
  \label{eq:currents}
  g\, j^{\mu\,\Lambda} = \frac{\delta {\cal L}_{\rm matter}}{\delta
  A_{\mu\,\Lambda}} \;,\qquad 
g\, j^\mu{}_\Lambda = \frac{\delta {\cal L}_{\rm matter}}{\delta
A_{\mu}{}^{\Lambda}} \;.
\end{equation}
Gauge invariance requires these currents to satisfy the following
constraints (subject to the matter field equations), 
\begin{equation}
  \label{eq:cons-currents}
  D_\mu\,j^\mu{}_M=0\,, \qquad \Theta^{\Lambda\alpha}\, j_{\mu\,\Lambda} =
  \Theta_\Lambda{}^\alpha \, j_{\mu}{}^\Lambda \,.  
\end{equation}
Equation~(\ref{eq:eom1}) is the duality equation that relates the
field strengths ${\cal H}_{\rho\sigma\,\Lambda}$ of the magnetic vector
fields to the electric field strengths via~(\ref{eq:def-G(H)}), at
least for the components projected by $\Theta^{\Lambda\alpha}$.
Equation~(\ref{eq:eom2}) relates the relevant tensor field strengths
(remember that the components of the tensor field other than 
$\Theta^{\Lambda\alpha} B_{\mu\nu\,\alpha}$ are not present in the
Lagrangian and thus do not lead to independent field equations) to the
gauge fields and the magnetic matter current.\footnote{
  In the presence of scalar fields, (\ref{eq:eom2}) takes the form of
  the duality equation that relates scalar and tensor fields. We shall
  return to this feature in the next section.}  
Equations~(\ref{eq:eom1}) and~(\ref{eq:eom2}) thus determine the field
strengths of the magnetic vectors and of the tensor fields,
respectively, in terms of the other fields. They do not play the role
of dynamical field equations, but together with the combined vector
and tensor gauge invariances they ensure that the number of
propagating degrees of freedom has not changed upon the introduction
of tensor and magnetic vector fields in the gauged theory.  We will
present a more explicit analysis of the degrees of freedom after
proper gauge fixing in the next section.  At $g=0$ both
(\ref{eq:def-G(H)}) and (\ref{eq:eom2}) are identically satisfied
which is consistent with the fact that in the ungauged theory the
tensor and magnetic vector fields drop from the Lagrangian.  Finally,
(\ref{eq:eom3}) via~(\ref{eq:def-G(H)}) constitutes the dynamical
equations of motion for $n$ vector fields.

\paragraph{Note added after publication:}
Using the Bianchi identity~(\ref{eq:vt-bianchi2}),
the equations of motion~(\ref{eq:eom1})--(\ref{eq:eom3})
may be recast into the manifestly covariant form
\bea
\Theta_{M}{}^{\alpha}\,({\cal H}_{\mu\nu}-{\cal G}_{\mu\nu})^{M}&=&0\;,
\qquad\;\;
\ft12\,\epsilon^{\mu\nu\rho\sigma}\,D_{\nu}\,{\cal G}_{\rho\sigma}{}^{M}
~=~ g\,\Omega^{MN} j^{\mu}{}_{N}
\;,
\eea
with the symplectic vector ${\cal G}_{\mu\nu}{}^{M}=
({\cal G}_{\mu\nu}{}^{\Lambda},{\cal G}_{\mu\nu\,\Lambda})
\equiv({\cal H}_{\mu\nu}{}^{\Lambda},{\cal G}_{\mu\nu\,\Lambda})$.

\section{Applications}
In this section we will illustrate a number of features of the general
results presented above. The universal Lagrangian~(\ref{eq:Ltotal})
presented in the last section combines tensor fields with electric and
magnetic vector fields.  We argued above that the total number of degrees of
freedom is independent of the embedding tensor, i.e.\ it remains
unchanged with respect to the ungauged theory owing to the fact that
magnetic vector and tensor fields appear with their own gauge
invariances and couple with a topological first-order kinetic term.
In concrete applications it is often useful to fix most of the gauge
invariances and eliminate the auxiliary fields in order to arrive at a
formulation in terms of only physical fields. The universal
Lagrangian~(\ref{eq:Ltotal}) offers various possibilities of gauge
fixing which lead to different effective Lagrangians that are related
by nonlocal field redefinitions and/or electric/magnetic duality.

Below, in subsection~\ref{sec:gf}, we present a general way of gauge
fixing by integrating out all the tensor fields from the Lagrangian.
This leads to an effective Lagrangian in terms of $n$ physical vector
fields and confirms the analysis of degrees of freedom given above.
The result can be interpreted as effecting an electric/magnetic
duality transformation directly at the level of the Lagrangian. In the
next subsection~\ref{sec:ag} we consider a particular class of abelian
gaugings generated by translational isometries which are often
relevant for the effective field theories that describe flux
compactifications.  We show that for these gaugings there is an
alternative way of gauge fixing which instead leads to a Lagrangian in
terms of electric vector fields and tensors upon eliminating some of
the scalar fields.  Finally, in subsection~\ref{sec:n2} we briefly
comment on the general results of this paper in the context of ${N}=2$
supergravity.

\subsection{Gauge fixing}
\label{sec:gf}
In this subsection we exhibit how the tensor fields can be integrated
out from the universal Lagrangian~(\ref{eq:Ltotal}) by choosing a
convenient basis for the embedding tensor. Upon further gauge fixing
of the remaining tensor gauge transformations this yields a Lagrangian
containing precisely $n$ physical vector fields.  We choose a basis of
the magnetic vector fields $A_\mu{}^\Lambda$ and the generators
$t_\alpha$ such that the rectangular matrix $\Theta^{\Lambda\alpha}$
decomposes into a square invertible submatrix $\Theta^{Ii}$ (with
inverse $(\Theta^{-1})_{iI}$), with all other submatrices
$\Theta^{Im}$, $\Theta^{Ui}$ and $\Theta^{Um}$ vanishing. Hence we
decomposed the $\mathrm{G}$-generators according to $t_\alpha\to
(t_i,t_m)$ and the magnetic vector fields $A_{\mu\Lambda}\to (A_{\mu I},A_{\mu
  U})$. Note that the decomposition of the generators and fields is
not yet completely fixed, as one can, for instance, redefine the $t_i$
by adding terms linear in the $t_m$. From (\ref{eq:quad}) we deduce
the following constraints on the remaining components of the embedding
tensor,
\begin{equation}
  \label{eq:constraint-basis}
  \Theta_I{}^m =0\,,\qquad \Theta^{Ii}\,\Theta_I{}^j =
  \Theta^{Ij}\,\Theta_I{}^i \,. 
\end{equation}
In this basis, equation~(\ref{eq:eom1}) takes the form
\begin{equation}
  \label{eq:BJ0}
  \left(\Theta_{I}{}^{i}+\Theta^{Ji}{\cal R}_{IJ}\right)
  B_{\mu\nu}{}^{I}
  -\ft12\varepsilon_{\mu\nu\rho\sigma} \Theta^{Ji}{\cal I}_{IJ}\,
  B^{\rho\sigma\,I} 
  = 2\, \Theta^{Ii}\, {\cal J}_{\mu\nu\,I}\;,
\end{equation}
with 
\begin{eqnarray}
  \label{eq:BJ-def}
  B_{\mu\nu}{}^{I} &=& g\,\Theta^{Ij}B_{\mu\nu\,j}\,, \nonumber\\  
  {\cal J}_{\mu\nu\,I} &=& {\cal F}_{\mu\nu\,I}-
  {\cal R}_{I\Lambda}{\cal F}_{\mu\nu}{}^{\Lambda} 
   +\ft12\epsilon_{\mu\nu\rho\sigma} {\cal I}_{I\Lambda}{\cal
   F}^{\rho\sigma\,\Lambda} + 
  \varepsilon_{\mu\nu\rho\sigma}{\cal O}^{\rho\sigma}{}_{I}\,,
\end{eqnarray}
Observe that no other tensor fields will appear in the Lagrangian by
virtue of (\ref{eq:constraint-basis}).  After some manipulation
(\ref{eq:BJ0}) gives rise to
\begin{equation}
  \label{eq:BJ}
  ({\cal I}+r\,{\cal I}^{-1}r)_{IJ}\,B_{\mu\nu}{}^{J} =
  \varepsilon_{\mu\nu\rho\sigma}\,{\cal J}^{\rho\sigma}{}_{I}+
  2\,(r\,{\cal I}^{-1}){}_{I}{}^{J}\,{\cal J}_{\mu\nu\,J} \;,
\end{equation}
with $r_{IJ}\equiv{\cal R}_{IJ}+(\Theta^{-1})_{iI}
\Theta^{\vphantom{-1}}_{J}{}^{i}$ a symmetric matrix and ${\cal
I}_{IK}({\cal I}^{-1}){}^{KJ} =\delta_{I}^{J}$. 
Substitution of this expression for $B_{\mu\nu}{}^I$ into the
Lagrangian~(\ref{eq:Ltotal}) leads to the following terms,
\begin{eqnarray}
  \label{eq:B-terms}
  \mathcal{L}_{\mathrm{B}} &=&
  \ft1{4}   [({\cal I}+r\,{\cal I}^{-1}r)^{-1}]^{IJ}\,\Big[
  \mathcal{J}_{\mu\nu I}\, \mathcal{J}^{\mu\nu}{}_J  - \ft12
  \varepsilon^{\mu\nu\rho\sigma} (r\,{\cal I}^{-1})_I{}^K\, 
  \mathcal{J}_{\mu\nu K}\,   \mathcal{J}_{\rho\sigma J} \Big]\;,
\end{eqnarray}
which should be added to the $B$-independent terms of the Lagrangian
(\ref{eq:Ltotal}), so that we are dealing with a Lagrangian that
depends on $2n$ vector fields. However, the magnetic vector fields
$A_{\mu \,U}$ are actually absent whereas the tensor gauge
transformations can be used to eliminate the electric vector fields
$A_\mu{}^I$ from the Lagrangian, Eventually one thus arrives at a
Lagrangian $\mathcal{L}_{\mathrm{V}}$ formulated in terms of $n$
vector fields $(A_{\mu}{}^{U}, A_{\mu\,I})$ carrying the $2n$ degrees
of freedom.

To see that the Lagrangian does not depend on the fields
$A_{\mu\,U}$, we first observe that neither $\mathcal{J}_{\mu\nu\,I}$
nor the $B$-independent terms in (\ref{eq:Ltotal}) contain the field
strengths $\mathcal{F}_{\mu\nu\,U}$. Hence in the abelian case the
absence of $A_{\mu\,U}$ is obvious. In the non-abelian case this is
less obvious. Although we know that $X^U = 0 = X_{(MN)}{}^U$, this
does not exclude that no upper indices $U$ will appear on the
generators. Fortunately the absence of $A_{\mu\,U}$ can be directly
inferred from the fact that (\ref{eq:eom2}) is proportional to
$\Theta^{\Lambda\alpha}$ which vanishes for $\Lambda=U$ (in
particular, the matter current $j^U=0$), so that we conclude that
$\delta{\cal L}_{\rm V} /\delta A_{\mu\,U}=0\,$. Here it is important
to realize that the tensor field equations (\ref{eq:eom1}) are
identically satisfied, so that the variations from $B_{\mu\nu}{}^I$ as
defined by (\ref{eq:BJ}) will not contribute. Note that we are
only dealing with the matter currents $j_{\mu\,U}$ and $j_\mu{}^I$ as
$j_\mu{}^U=0$ and $j_{\mu\,I} = (\Theta^{-1})_{iI} \Theta_J{}^i\;
j_\mu{}^J$.

We now combine (\ref{eq:Ltotal}) and (\ref{eq:B-terms}) to find the
new Lagrangian ${\cal L}_{{\rm V}}$. For simplicity we evaluate this
Lagrangian in the abelian case without 
moment couplings, so that the only quantities involved are the abelian
field strengths $\mathcal{F}_{\mu\nu I}$ and
$\mathcal{F}_{\mu\nu}{}^U$. The result takes the following form, 
\begin{eqnarray}
\label{eq:new-Ltotal}
  {\cal L}_{{\rm V}} &=&
  \ft14 \Big[ \hat{\cal I}^{IJ}\,\mathcal{F}_{\mu\nu I} \,
   \mathcal{F}^{\mu\nu}{}_J + \hat{\cal
   I}_{UV}\,\mathcal{F}_{\mu\nu}{}^U\, 
   \mathcal{F}^{\mu\nu V} + 2\,\hat{\cal I}^I{}_U
   \,\mathcal{F}_{\mu\nu I}\, \mathcal{F}^{\mu\nu U} \Big]    
  \nonumber\\[.9ex]
  &&{}
  + \ft18 \varepsilon^{\mu\nu\rho\sigma} 
  \Big[ \hat{\cal R}^{IJ}\,\mathcal{F}_{\mu\nu I} \,
   \mathcal{F}_{\rho\sigma J} + \hat{\cal
   R}_{UV}\,\mathcal{F}_{\mu\nu}{}^U\, \mathcal{F}_{\rho\sigma}{}^V 
   + 2\,\hat{\cal R}^I{}_U    \,\mathcal{F}_{\mu\nu I} \,
   \mathcal{F}_{\rho\sigma}{}^U \Big]    \;,
\end{eqnarray}
where
\begin{eqnarray}
  \label{eq:hat-I-R}
  \hat\mathcal{I}^{IJ} &=& [({\cal I}+r\,{\cal I}^{-1}r)^{-1}]^{IJ}\;,  
   \nonumber\\
  \hat\mathcal{I}_{UV} &=&  \mathcal{I}_{UV}  \nonumber \\
  &&{} 
  +   [({\cal I}+r\,{\cal I}^{-1}r)^{-1}]^{IJ}  
  \Big[\mathcal{R}_{UI}\,\mathcal{R}_{JV} 
       - \mathcal{I}_{UI}\, \mathcal{I}_{JV} 
  -  2\,(r\mathcal{I}^{-1})_J{}^K\,\mathcal{R}_{I(U} \,\mathcal{I}_{V)K}
  \Big]\;,
   \nonumber\\
  \hat\mathcal{I}^I{}_U &=&   [({\cal I}+r\,{\cal I}^{-1}r)^{-1}]^{IJ}
   \Big[ - \mathcal{R}_{JU} + (r\mathcal{I}^{-1})_J{}^K\,\mathcal{I}_{KU}
   \Big] \;, \nonumber\\
  \hat\mathcal{R}^{IJ} &=&  {}-  [({\cal I}+r\,{\cal I}^{-1}r)^{-1}]^{IK}
  \, (r\mathcal{I}^{-1})_K{}^J \;, 
   \nonumber\\
  \hat\mathcal{R}_{UV} &=&    \mathcal{R}_{UV} \nonumber \\
  &&{} 
  +   [({\cal I}+r\,{\cal I}^{-1}r)^{-1}]^{IJ} 
  \Big[(-\mathcal{R}_{IU}\,\mathcal{R}_{VK} 
       + \mathcal{I}_{IU}\, \mathcal{I}_{VK})
         (r\mathcal{I}^{-1})_J{}^K 
  - 2\,\mathcal{R}_{I(U} \, \mathcal{I}_{V)J} \Big]\;, 
   \nonumber\\
  \hat\mathcal{R}^I{}_U &=&   [({\cal I}+r\,{\cal I}^{-1}r)^{-1}]^{IJ}
  \Big[\mathcal{I}_{JU} + (r\mathcal{I}^{-1})_J{}^K \,\mathcal{R}_{KU}
  \Big] \;.   
\end{eqnarray}
We note that (in contrast to the situation in odd dimensions) this
gauge-fixed Lagrangian allows a smooth limit $g\rightarrow0$. At
$g=0$, however, this does not bring back the original
Lagrangian~(\ref{eq:quadratic-L}) but rather one related to it by
electric/magnetic duality. To see this one first performs a shift of
the generalized theta angle, $\mathcal{R}_{IJ}  \to \mathcal{R}_{IJ}+
(\Theta^{-1})_{iI} \Theta^{\vphantom{-1}}_{J}{}^{i}= r_{IJ}$, followed
by a second duality transformation where (\ref{eq:em-duality}) is the
unit matrix in the subspace carrying indices $U,V$, whereas in
the subspace carrying the indices $I,J$ it is an off-diagonal
transformation with $W=-Z=\mathbf{1}$ (in other words, the typical
strong-weak coupling duality). Hence in this formalism one is able to
perform duality transformations at the level of the local Lagrangian. 

\subsection{Abelian gaugings}
\label{sec:ag}
In many situations one is dealing with a
group $\mathrm{G}$ of symmetries of the ungauged theory that
factorizes into two groups, one of which acts exclusively on the 
matter fields. This situation
is, for instance, relevant for abelian gaugings, where the vector
fields transform in a trivial representation and the matter fields
transform in a non-trivial representation of the (abelian) gauge
group. In that case the gauge group can be embedded into a group that
acts exclusively on the matter fields. Many supersymmetric models
show this feature. 

Assuming that the gauge group will be embedded into a rigid invariance
group that is decomposable into
$\mathrm{G}_{\mathrm{V}}\times\mathrm{G}_{\mathrm{M}}$, where
$\mathrm{G}_{\mathrm{M}}$ acts exclusively on the matter fields, 
we decompose the generators accordingly into two mutually commuting sets:
$\{t_{\alpha}\}=\{t_{A}\} \oplus \{t_{a}\}$, where only the generators
$t_{A}$ induce a nontrivial action on the vector fields. 
The latter implies that the $(t_a)_M{}^N$ vanish as these generators 
act exclusively in the matter sector. Obviously we are dealing with two
sets of structure constants, $f_{AB}{}^C$ and $f_{ab}{}^c$. The
embedding tensor $\Theta_M{}^\alpha$ decomposes into $\Theta_M{}^A$
and $\Theta_M{}^a$, which define the gauge group generators $X_M=
\Theta_M{}^At_A +\Theta_M{}^a t_a$. The quadratic
constraint (\ref{eq:clos}) then decomposes into two separate equations,
\begin{eqnarray}
 f_{AB}{}^{C}\, \Theta_{M}{}^{A}\,\Theta_{N}{}^{B}
+(t_{A})_{N}{}^{P}\,\Theta_{M}{}^{A}\Theta_{P}{}^{C} &=&0 \;, 
\label{eq:quad2b} \\[1ex]
  f_{ab}{}^{c}\, \Theta_{M}{}^{a}\,\Theta_{N}{}^{b}
  +(t_{A})_{N}{}^{P}\,\Theta_{M}{}^{A}\Theta_{P}{}^{c} &=&0 \;.
\label{eq:quad2a}
\end{eqnarray}
The second quadratic constraint (\ref{eq:constraint-eq}) leads to an
additional condition (see also, the comment below
(\ref{eq:constraint-eq})), 
\begin{equation}
  \label{eq:quadratic-matter}
  \Theta^{\Lambda[a} \Theta_\Lambda{}^{b]} = 0\;.
\end{equation}

For abelian gaugings we have $\Theta_M{}^A=0$ and the
commutativity of the matter charges is ensured by
(\ref{eq:quad2a}). The vector/tensor 
Lagrangian for abelian gaugings takes a rather simple form,
\begin{eqnarray}
\label{eq:Ltotal-abelian}
  {\cal L}_{{\rm VT}} &=&
\ft14 \, {\cal I}_{\Lambda\Sigma}\,\mathcal{H}_{\mu\nu}{}^{\Lambda} 
\mathcal{H}^{\mu\nu\,\Sigma} 
+\ft 18 {\cal R}_{\Lambda\Sigma}\;\varepsilon^{\mu\nu\rho\sigma} 
\mathcal{H}_{\mu\nu}{}^{\Lambda} 
\mathcal{H}_{\rho\sigma}{}^{\Sigma}  
  +\mathcal{H}_{\mu\nu}{}^{\Lambda}
  \mathcal{O}^{\mu\nu}{}\!_\Lambda \nonumber \\[.9ex]
  && {}
  +  \ft12 [\mathcal{I}^{-1}]^{\Lambda\Sigma} \, \mathcal{O}_{\mu\nu\,\Lambda}
  \mathcal{O}^{\mu\nu}{}\!_\Sigma
  \nonumber\\[.9ex]
&&{}
-\ft14g\, \varepsilon^{\mu\nu\rho\sigma}\,
\Theta^{\Lambda a}\,B_{\mu\nu\,a} \,
\partial_{\rho} A_{\sigma\,\Lambda} 
 + \ft1{32} g^2\, 
   \Theta^{\Lambda\,a}\Theta_{\Lambda}{}^{b}\, \varepsilon^{\mu\nu\rho\sigma}\,
  B_{\mu\nu\,a} \,B_{\rho\sigma\,b} \;,
\end{eqnarray}
where 
\begin{equation}
  \label{eq:abelian-H}
  \mathcal{H}_{\mu\nu}{}^{\Lambda}= 2\,\partial_{[\mu}
  A_{\nu]}{}^{\Lambda} + \ft12g\, 
  \Theta^{{\Lambda}\,a} B_{\mu\nu\,a} \,.
\end{equation}

A particular example of abelian gaugings concerns the case of a
nonlinear sigma model with gauged translational isometries of its 
scalar target space.  Such gaugings for instance appear in Calabi-Yau
(or half-flat manifold) compactifications in the presence of
background fluxes~\cite{Taylor:1999ii,DallAgata:2001zh,Louis:2002ny,
  Gurrieri:2002wz,Gurrieri:2002iw,DAuria:2004wd}.  Let us thus
consider a scalar target space parametrized by scalar fields
$\{\phi^{a},q^{i}\}$ whose metric $G_{mn}$ does not depend on the
subset $\{q^{i}\}$ of scalar fields such that the shifts
$q^{i}\rightarrow q^{i}+c^{i}$ constitute a set of abelian isometries.
A gauging of these isometries is encoded in an embedding
tensor $\Theta_{M}{}^{i}=(\Theta_{\Lambda}{}^{i},\Theta^{\Lambda\,i})$
subject to~(\ref{eq:quadratic-matter}).  It induces the covariant
derivatives
\begin{equation}
D_{\mu}q^{i}=  \partial_{\mu}q^{i}-
  g A_{\mu}{}^{\Lambda}\Theta_{\Lambda}{}^{i}  -g
A_{\mu\,\Lambda}\Theta^{\Lambda\,i}  \;.  
\end{equation}
The magnetic vector fields $\Theta^{\Lambda\,I}A_{\mu\,\Lambda}$ can
then be integrated out using the equations of motion~(\ref{eq:eom2}), 
\begin{equation}
  \varepsilon^{\mu\nu\rho\sigma}\,
  \partial_{\nu}B_{\rho\sigma\,i} ~\propto~
  G_{ia}(\phi)\,\partial^{\mu}\phi^{a} +G_{ij}(\phi)\,\Big(
  \partial^{\mu}q^{j} -gA^{\mu}{}^{\Lambda}\Theta_{\Lambda}{}^{j}
  -gA^{\mu}{}_{\Lambda}\Theta^{\Lambda\,j}\Big) \;.
\end{equation}
This shows that the topological term~(\ref{eq:Ltop}) eventually gives
rise to a topological coupling
$\varepsilon^{\mu\nu\rho\sigma}\,\Theta_{\Lambda}{}^{i}
B_{\rho\sigma\,i}\,\partial_{\mu} A_{\nu}{}^{\Lambda}$ between tensor
and electric vector fields as well as to a kinetic term
$(G^{-1}){}^{ij}(\phi)\;\partial_{[\mu}B_{\nu\rho]\,i}\;
\partial^{[\mu}B^{\nu\rho]}{}_{j}$ for the tensor fields.  This leads
to a Lagrangian whose physical fields comprise tensor and electric
vector fields, which reproduces the results of
\cite{Louis:2002ny,Sommovigo:2004vj, DAuria:2004yi}
Alternatively, following the gauge fixing procedure described in the
previous subsection, leads instead to a Lagrangian expressed exclusively
in terms of (electric and magnetic) vector fields.  The general
formalism presented here allows rather straightforward generalizations 
involving the gauging of nonabelian isometries in the presence of
tensor fields.  Integrating out scalar and magnetic vector fields in
the nonabelian case will presumably lead to the non-polynomial
interactions of tensor fields captured by (extensions of) the
Freedman-Townsend models~\cite{Freedman:1980us, Henneaux:1997ha,
  Theis:2004pa}.
Other applications or generalizations may, for instance, 
involve M-theory compactifications on twisted tori~\cite{DFT1,DFT2}.

\subsection{$N=2$ supersymmetry}
\label{sec:n2}
As a final topic we briefly discuss gaugings of $N=2$ supergravity,
where the scalar target space is a direct product of a
special-K\"ahler and a quaternion-K\"ahler manifold whose coordinates
we denote by complex fields $z^{i}$ and real $q^{u}$, respectively.
The isometry group factors into the direct product ${\rm G}_{{\rm
    SK}}\times {\rm G}_{{\rm Q}}$.  Only the generators of ${\rm
  G}_{{\rm SK}}$ induce a nontrivial action on the vector fields. This
is a special case of the situation described in the beginning of the
previous subsection. Accordingly, we label the generators of the isometry
groups as $\{t_{\alpha}\}=\{t_{A}\} \oplus \{t_{a}\}$.  A gauging is
encoded in an embedding tensor $\Theta_M{}^\alpha=(\Theta_M{}^A,
\Theta_M{}^a)$, subject 
to the constraints (\ref{eq:quad2a}) and (\ref{eq:quadratic-matter}).
The kinetic term for the scalar fields is described by a nonlinear
sigma-model 
\begin{equation}
  \label{eq:N2-kinetic}
   {\cal L}_{\rm kin} = -\ft12\,g_{i\bar\jmath}\,
   D_{\mu}z^{i}\,D^{\mu}\overline z{}^{\bar\jmath}\, -\ft12\,h_{uv}\, 
   D_{\mu}q^{u}\,D^{\mu}q^{v} \;,  
  \end{equation}
where $g_{i\bar\jmath}$ and $h_{uv}$ denote the metrics on the
special-K\"ahler and the quaternion-K\"ahler manifold, respectively,
and the covariant derivatives,
\begin{eqnarray}
  \label{eq:cov-der-N=2}
D_{\mu}z^{i} &=& \partial_{\mu} z^{i} -g\, \Theta_{M}{}^{A}
A_{\mu}{}^{M} k^{i}{}\!_{A} \;,
\nonumber\\
D_{\mu}q^{u} &=& \partial_{\mu} q^{u} -g\, \Theta_{M}{}^{a}
A_{\mu}{}^{M} k^{u}{}\!_{a} \;, 
\end{eqnarray}
are written in terms of the embedding tensor and the corresponding
Killing vector fields $k^{i}{}\!_{A}$ and $k^{u}{}\!_{a}$.  For gaugings
that involve only electric vector fields $N=2$ supersymmetry requires a
scalar potential~\cite{DAuria:1990fj,DAuria:2001kv,deWit:2001bk}, 
which can be written as follows, 
\begin{eqnarray}
  \label{eq:potential}
  V&=& L^{\Lambda} \bar L^{\Sigma}\,
  \Big(\Theta_{\Lambda}{}^{A}\Theta_{\Sigma}{}^{B}\,
  g_{i\bar\jmath}\,k^{i}{}\!_{A}\,k^{\bar\jmath}{}\!_{B} +
  4\,\Theta_{\Lambda}{}^{a}\Theta_{\Sigma}{}^{b}\, h_{uv}\,
  k^{u}{}\!_{a}\,k^{v}{}\!_{b} \Big) 
\nonumber\\[1ex]
&& {} + \vec{\cal P}_{a}\cdot\vec{\cal P}_{b}\;
\Theta_{\Lambda}{}^{a}\Theta_{\Sigma}{}^{b} \, \Big(g^{i\bar\jmath}
f_{i}{}^{\Lambda} \bar{f}_{\bar\jmath}{}^{\Sigma}
-3\,L^{\Lambda}\bar{L}^{\Sigma}\Big) \;.  
\end{eqnarray}
Here, $L^{\Lambda}$
denotes the upper half of the symplectic section $L^{M}=
(L^{\Lambda},M_{\Lambda})\equiv e^{{\cal
    K}/2}(X^{\Lambda},F_{\Lambda})$ on the special K\"ahler manifold
with K\"ahler potential ${\cal K}$, and $ f_{i}{}^{\Lambda} \equiv
(\partial_{i}\!+\!\ft12\partial_{i}{\cal K})\, L^{\Lambda}$ denotes
its K\"ahler covariant derivative; the $\mathrm{Sp}(1)$ vectors 
$\vec{\cal P}_{a}$ are the quaternion-K\"ahler moment maps 
associated with the Killing vectors $k^u{}_a$. 

It is now straightforward to generalize this expression to a situation
where both electric and magnetic vector fields are involved in the
gauging. Here we recall that the potential arises as a supersymmetric
completion associated with the gauging. However, the electric-magnetic
duality plays only an ancillary role in this sector, as is known, for
instance, from the gaugings in maximal supergravity theories. There it
was demonstrated that the so-called $T$-tensors are directly
expressible in terms of the embedding tensor without the necessity of
making a distinction between magnetic and electric components.  In
fact electric and magnetic components of the embedding tensor 
can only be identified by referring to the kinetic terms of the
vector fields. Hence, the embedding tensor and the $T$-tensor, and
thus the potential (which is quadratic in the $T$-tensor) is
insensitive to these features and does not change under vector-tensor
and vector-vector dualities \cite{deWit:2004nw,deWit:2002vt,dWST4}.

With the above observations in mind, we may thus write the full scalar
potential as a symplectically covariant expression (treating the
embedding tensors as a spurionic quantity),  
\begin{eqnarray}
  \label{eq:e-m-potential}
   V&=& L^{M} \bar L^{N}\,
   \Big(
   \Theta_{M}{}^{A}\Theta_{N}{}^{B}\,
   g_{i\bar\jmath}\,k^{i}{}\!_{A}k^{\bar\jmath}{}\!_{B} +
   4\,\Theta_{M}{}^{a}\Theta_{N}{}^{b}\, h_{uv}\,
   k^{u}{}\!_{a}k^{v}{}\!_{b}  \Big)
   \nonumber\\[1ex]
   && {}
   +\vec{\cal P}_{a}\cdot \vec{\cal P}_{b}\;
   \Theta_{M}{}^{a}\Theta_{N}{}^{b} \, \Big(g^{i\bar\jmath}
   f_{i}{}^{M}  \bar{f}_{\bar\jmath}{}^{N} -3\,
   L^{M}\bar{L}^{N}\Big)\;. 
\end{eqnarray}
For the abelian case, where $\Theta_{M}{}^{A}=0$, this expression
coincides with the one presented long ago in~\cite{Michelson:1996pn}.
Of course, a full supersymmetric derivation requires to cast the
results of this paper in a supersymmetric context. 

\vspace{8mm}
\noindent
{\bf Acknowledgement}\\
\noindent
We wish to thank L.~Andrianopoli, F.~Saueressig, S.~Vandoren and
M.~Weidner for helpful discussions.  H.S. and M.T. thank the CERN
theory division for hospitality while this work was started; B.d.W.
thanks the Benasque Center for Science, where the work was completed
at the String Theory Workshop.  This work is partly supported by EU
contracts MRTN-CT-2004-005104, MRTN-CT-2004-503369, and
MRTN-CT-2004-512194, the INTAS contract 03-51-6346, and the DFG grant
SA 1336/1-1.

\bigskip

\providecommand{\href}[2]{#2}
\begingroup\raggedright\endgroup
\end{document}